# Characterization of AlGaAs/GeSn heterojunction band alignment via X-ray photoelectron spectroscopy

Yang Liu[a,1], Jiarui Gong[a,1], Sudip Acharya[b,1], Yiran Li[a], Alireza Abrand[c],

Justin M. Rudie[b], Jie Zhou[a], Yi Lu[a], Haris Naeem Abbasi[a], Daniel Vincent[a],

Samuel Haessly[a], Tsung-Han Tsai[a], Parsian K. Mohseni[c,*], Shui-Qing Yu[b,*], and

Zhenqiang Ma[a,*]

[a]Department of Electrical and Computer Engineering, University of Wisconsin-Madison, Madison, WI 53706, USA

Email: mazq@engr.wisc.edu

[b]Department of Electrical Engineering, University of Arkansas, Fayetteville, AR 72701 USA

Email: syu@uark.edu

[c]Microsystem Engineering, Rochester Institute of Technology, Rochester, NY 16423, USA

Email: pkmohseni@rit.edu

[1] Yang Liu, Jiarui Gong, and Sudip Acharya contributed equally to this work

**Abstract**

GeSn-based SWIR lasers featuring imaging, sensing, and communications has gained dynamic development recently. However, the existing SiGeSn/GeSn double heterostructure lacks adequate electron confinement and is insufficient for room temperature lasing. The recently demonstrated semiconductor grafting technique provides a viable approach towards AlGaAs/GeSn p-i-n heterojunctions with better electron confinement and high-quality interfaces, promising for room temperature electrically pumped GeSn laser devices. Therefore, understanding and quantitatively characterizing the band alignment in this grafted heterojunction is crucial. In this study, we explore the band alignment in the grafted monocrystalline $Al_{0.3}Ga_{0.7}As$ /$Ge_{0.853}Sn_{0.147}$ p-i-n heterojunction. We determined the bandgap values of AlGaAs and GeSn to be 1.81 eV and 0.434 eV by photoluminescence measurements, respectively. We further conducted X-ray photoelectron spectroscopy measurements and extracted a valence band offset of 0.19 eV and a conduction band offset of 1.186 eV. A Type-I band alignment was confirmed which effectively confining electrons at the AlGaAs/GeSn interface. This study improves our understanding of the interfacial band structure in grafted AlGaAs/GeSn heterostructure, providing experimental evidence of the Type-I band alignment between AlGaAs and GeSn, and paving the way for their application in laser technologies.

*Keywords:* **Band alignment, X-ray photoelectron spectroscopy, Germinum Tin, Semiconductor grafting, Electron confinement**

## 1. Introduction

Over the past decades, shortwave infrared (SWIR) light sources with wavelengths ranging from 1.6 to 3 µm have gained significant interest due to their unique interactions with various materials and penetration capabilities in different atmospheric conditions[1]. Diverse applications of the SWIR light sources have been proposed in fields such as aerial-based imaging systems[2-4], industrial inspections[5-7], night vision technologies[8], biomedical imaging and diagnostics[9,10], chemical analysis spectroscopy[11], and surveillance systems[12]. A variety of III-V compounds with narrow-bandgap materials, such as InGaAsSb[13], InP[14], and InAs[15], have been utilized to develop SWIR lasers. Although III-V lasers heterogeneously integrated on silicon platforms have achieved high-performance, their widespread adoption is hindered by high costs and scalability challenges, arising from complex fabrication processes[16,17] and difficulties in material growth[16]. These challenges limit the development of CMOS-compatible SWIR light sources on silicon for monolithic optoelectronic integrated circuits (OEICs).

Group IV materials, particularly GeSn alloys, show considerable promise in overcoming these obstacles due to their potential for a direct bandgap transition[18] and CMOS-compatibility[19]. With an adequate Sn concentration, the energy in the Γ-valley can drop below that in the L-valley, transitioning GeSn from an indirect to a direct bandgap material[20]. This transition is typically observed when the Sn content is between 6% and 10%. The first GeSn laser, reported by Wirths et al. in 2015, operated under optical pumping at temperatures up to 90K[21]. Subsequent developments have led to significant enhancements in GeSn optically pumped lasers, with room-temperature lasing achieved in microdisks[22] and bridges[23] due to the tensile-strain enhanced energy splitting between the L-valley and Γ-valley.

Electrically pumped GeSn lasers, which offer better integration potential, are highly desirable. Promising structures such as the SiGeSn/GeSn/SiGeSn double heterostructure[24] and the GeSn/SiGeSn multi-quantum-well structure [25,26]were designed and simulated in 2010. The first electrically injected Fabry-Perot GeSn laser demonstrated in 2020 used a SiGeSn/GeSn/GeSn double heterostructure and operated in pulsed mode with a maximum lasing temperature of 100K[27]. In this structure, SiGeSn is lattice-matched and exhibits Type-I band alignment with the GeSn active layer, providing carrier injection and optical confinement. At the cap/active region interface, $Si_{0.03}Ge_{0.89}Sn_{0.08}$ creates a 114 meV barrier height to $Ge_{0.89}Sn_{0.11}$, which enhances electron confinement and optical confinement due to its lower refractive index[28]. So far, the same double heterostructure has successfully achieved lasing at 140K by optimizing the layer design[29]. However, while such double heterostructure (DHS) confinement is effective at low temperatures, it becomes insufficient near room temperature due to carrier leakage excited by thermal energy[30]. Therefore, developing DHS with enhanced electron confinement barriers, while maintaining good optical confinement, is critical for improving device performance at higher temperatures.

$Al_{0.3}Ga_{0.7}As$ is pivotal in the advancement of laser technologies due to its higher band gap, higher conduction band energy, and relatively lower refractive index (~ 3.4) compared to GeSn (~ 4.2) and SiGeSn (~ 4.0)[31] [32]. These properties enable $Al_{0.3}Ga_{0.7}As$ to form Type-I band alignments with GeSn, facilitating robust electrical and optical confinement within heterostructures. Compared to using SiGeSn as cap material in lattice-matched SiGeSn/GeSn heterostructure, AlGaAs offers better electrical and optical confinement. However, growing high-quality III-V compounds like AlGaAs on Group IV materials presents challenges such as dislocations and defects that can trap carriers or increase non-radiative recombination, thereby undermining the reliability and efficiency of the devices[33,34]. Additionally, methods such as wafer bonding struggle with thermal mismatches and the potential for wafer cracking[35,36].

Semiconductor grafting is a cutting-edge technique that offers significant advancements in the development of heterostructures from dissimilar materials, overcoming issues related to lattice mismatches and thermal discrepancies[37]. This innovative approach employs an ultrathin dielectric layer, like an ultrathin oxide (UO), serving dual functions as a double-side passivation layer and a quantum tunnelling medium at the interface between two single-crystalline semiconductors. This method not only reduces the density of states at the heterointerface but also ensures an abrupt transition, enhancing the mechanical, optical, and electronic properties of the grafted heterostructure[38-42]. Such method has been proven to be indispensable in our previous work[43-45].

In this study, a grafted monocrystalline $Al_{0.3}Ga_{0.7}As$ / $Ge_{0.853}Sn_{0.147}$ heterojunction was successfully fabricated following the grafting approach. We employed X-ray photoelectron spectroscopy (XPS) to analyse the band alignment in the grafted $Al_{0.3}Ga_{0.7}As$ / $Ge_{0.853}Sn_{0.147}$ heterojunction. We collected core level peaks and valence band spectra for $Al_{0.3}Ga_{0.7}As$, $Ge_{0.853}Sn_{0.147}$, and the grafted $Al_{0.3}Ga_{0.7}As$ / $Ge_{0.853}Sn_{0.147}$ heterostructure (hereafter referred to as AlGaAs/GeSn). Using photoluminescence (PL) data to determine the bandgaps of $Al_{0.3}Ga_{0.7}As$ and $Ge_{0.853}Sn_{0.147}$, we then applied the core level method, as proposed by Kraut et al.[46] for calculation. The valence band offset (VBO) is -0.19 eV and conduction band offset (CBO) is 1.186 eV, forming a Type-I band alignment between the two materials. The determined AlGaAs/GeSn Type-I band alignment supports our hypothesis, providing better electrical confinement than SiGeSn/GeSn and Ge/GeSn heterostructures, and paves the way for future applications in laser technology.

## 2. Experiment

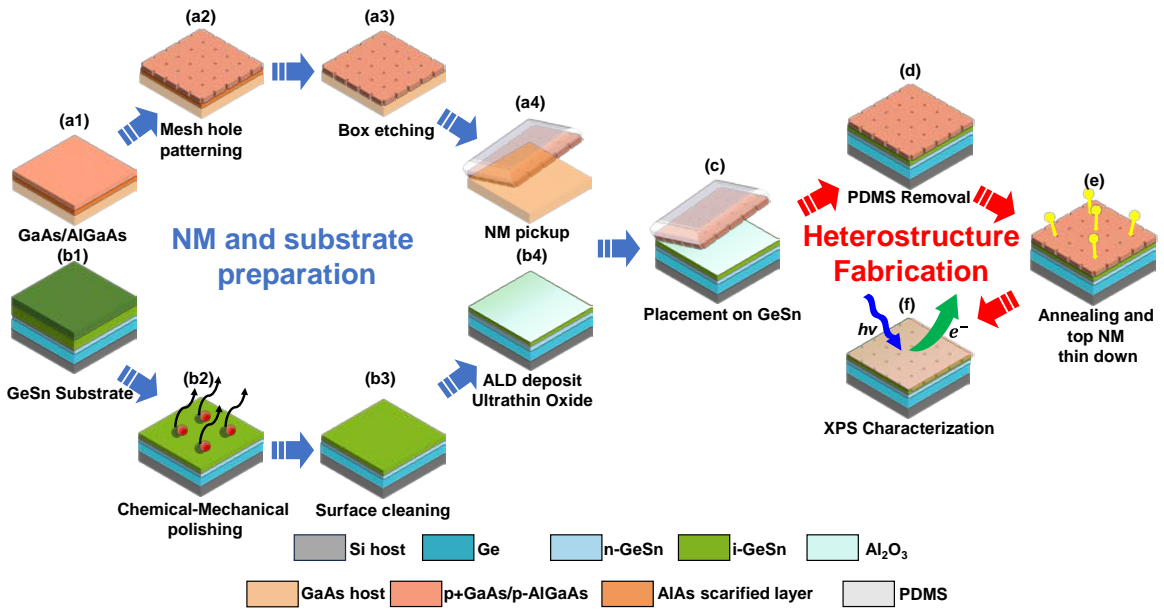

Fig.1. Schematic illustration of the fabrication procedure of the grafted monocrystalline AlGaAs/ Ge$_{0.853}$Sn$_{0.147}$ heterostructure sample for the XPS characterization. (a1) GaAs/AlGaAs/AlAs structure grown on a semi-insulating (SI) GaAs substrate; (a2) Standard mesh pattern lithography and holes etching using ICP-RIE etcher on the GaAs/AlGaAs structure; (a3) Box etching sacrificial layer using hydrofluoric acid; (a4) p$^+$-GaAs/ p$^-$-AlGaAs nanomembrane (NM) pickup by polydimethylsiloxane(PDMS) stamp; (b1) GeSn substrate p-Si$_{0.03}$Ge$_{0.89}$Sn$_{0.08}$/i- Ge$_{0.853}$Sn$_{0.147}$/n$^+$-GeSn/Ge buffer layer grown on the Si substrate; (b2) Chemical-mechanical polishing the GeSn substrate; (b3) GeSn surface cleaning; (b4) ALD coating of the 5 cycles Al$_2$O$_3$ on the GeSn surface. (c) Transfer of the released GaAs/AlGaAs NM to the GeSn substrate; (d) Removal of the PDMS from the heterostructure; (e)rapid thermal annealing process at 200 °C for 30 mins and GaAs/AlGaAs layer thinning down; (f) XPS characterization on the AlGaAs/GeSn interface.

The procedure of the AlGaAs/GeSn heterostructure started with the preparation of the AlGaAs nanomembrane (NM). An epitaxially grown wafer in Fig. 1(a1) consisting of a 100 nm $1\times10^{19}$ cm$^{-3}$ p$^+$-GaAs sitting on the top of the $1\times10^{18}$ cm$^{-3}$ p$^-$-AlGaAs 500 nm active layer, and a 400 nm AlAs sacrificial layer grown on a semi-insulating GaAs substrate is used to prepare the NM. The $9\times9$ µm$^2$ mesh holes patterning was then created by standard lithography, followed by a dry etching process using a Chlorine-based (BCl$_3$:10 sccm, Ar: 5 sccm) ICP plasma etcher to expose the underlying AlAs sacrificial layer, as shown in the Fig.1(a2)[47]. The etched epi was soaked in diluted hydrofluoric acid (HF: H$_2$O = 1:250) to remove the AlAs layer and fully released the p$^+$-GaAs/ p$^-$-AlGaAs standing on the GaAs substrate in Fig. 1(a3). Then the released p$^+$-GaAs/ p$^-$-AlGaAs went through a lift-off procedure using polydimethylsiloxane (PDMS) stamp following by an infiltration into Tetramethylammonium hydroxide (TMAH) based developer to remove the residual AlF$_3$ and Al(H$_2$O)n$^{3+}$ on the NM back side, as shown in Fig. 1(a4).

Meanwhile, a gentle chemical mechanical polishing (CMP) was performed on the GeSn epi (Fig. 1(b1)) layer, which contains 180 nm $1\times10^{18}$ cm$^{-3}$ p-Si$_{0.03}$Ge$_{0.89}$Sn$_{0.08}$, a 540 nm intrinsic Ge$_{0.853}$Sn$_{0.147}$, a 1.1 µm grading $1\times10^{19}$ cm$^{-3}$ n$^+$-GeSn/Ge buffer layer grown on the Si substrate. The secondary ion mass spectrometry (SIMS) profile of the original GeSn substrate before CMP is shown in Fig. 2(g), indicating good agreement with our desired growth structure. The CMP process shown in Fig. 1(b2) went through 165 s to remove the p-Si$_{0.03}$Ge$_{0.89}$Sn$_{0.08}$ cap layer and achieve a smooth surface (Rq= 0.42 nm, measured by an Atomic Force Microscope (AFM), see below) using ethylene glycol mixing with hydrogen peroxide. The remaining intrinsic GeSn layer is approximately 300 nm, calculated by the measured polishing rate at 150nm per minute. The polished substrate was then soaked in diluted hydrofluoric acid with DI water (1:20) for 1 minute to ensure the removal of polishing residues, as shown in Fig. 1(b3). The cleaned GeSn epi was immediately loaded into an Ultratech atomic layer deposition (ALD) system, depositing 5 cycles Al$_2$O$_3$ (0.11 nm/cy) on the surface at 200 ºC, as shown in Fig. 1(b4).

The AlGaAs NM in Fig. 1(a1) was then transfer-printed to the $Al_2O_3$-coated epitaxial GeSn destination substrate, as shown in Fig. 1(c). The schematic illustration of the grafted AlGaAs/GeSn heterojunction is shown in the Fig. 2(c). After removing the PDMS stamp in the Fig. 1(d), a rapid thermal annealing (RTA) process was conducted at 200 °C for 30 min to form a robust chemical bonding between the AlGaAs and $Ge_{0.853}Sn_{0.147}$ substrate (Fig. 1e). The annealing temperature and duration were chosen based on our experiments that consider the possible Sn segregation of the GeSn epi substrate. The NM was then thinned from 600 nm down to ~ 10 nm while maintaining its single crystallinity by a precisely controlled dry etching process using the same PlasmaTherm ICP-RIE etcher as mentioned in the previous step. A completed AlGaAs/GeSn heterostructure was then ready for the subsequent XPS characterization. After the ICP thinning process on AlGaAs, the sample was immediately stored in a nitrogen ambient container to minimize the air exposure time, and then was transferred to XPS chamber for characterizations.

To construct the AlGaAs/GeSn band diagram, the XPS measurements were performed on the pristine GeSn substrate (Fig. 1(b1)), the 400 nm AlGaAs sample after dry etching removal of the top GaAs layer, and the grafted AlGaAs/Ge heterostructure sample (Fig. 1(d)) through a Thermo Scientific K Alpha X-ray Photoelectron Spectrometer (XPS) with an Al $K_\alpha$ X-ray source ($hv$ = 1486.6 eV). The instrument was calibrated by Au $4f_{7/2}$ peak at 84.0 eV. The following settings were applied to the spectrometer: 10 eV pass energy, 400 μm spot size, up to 10 s dwell time, and 0.02 eV step size. Details regarding sample handling can be found in our previous work[44]. It is worthwhile to note that a 200 meV ion-beam etcher was applied on the surface during measurement to thin down the NM to around 5 nm, removing the possible oxidation during load lock and transfer, thereby enhancing the collection of interfacial spectral data.

## 3. Results and discussion

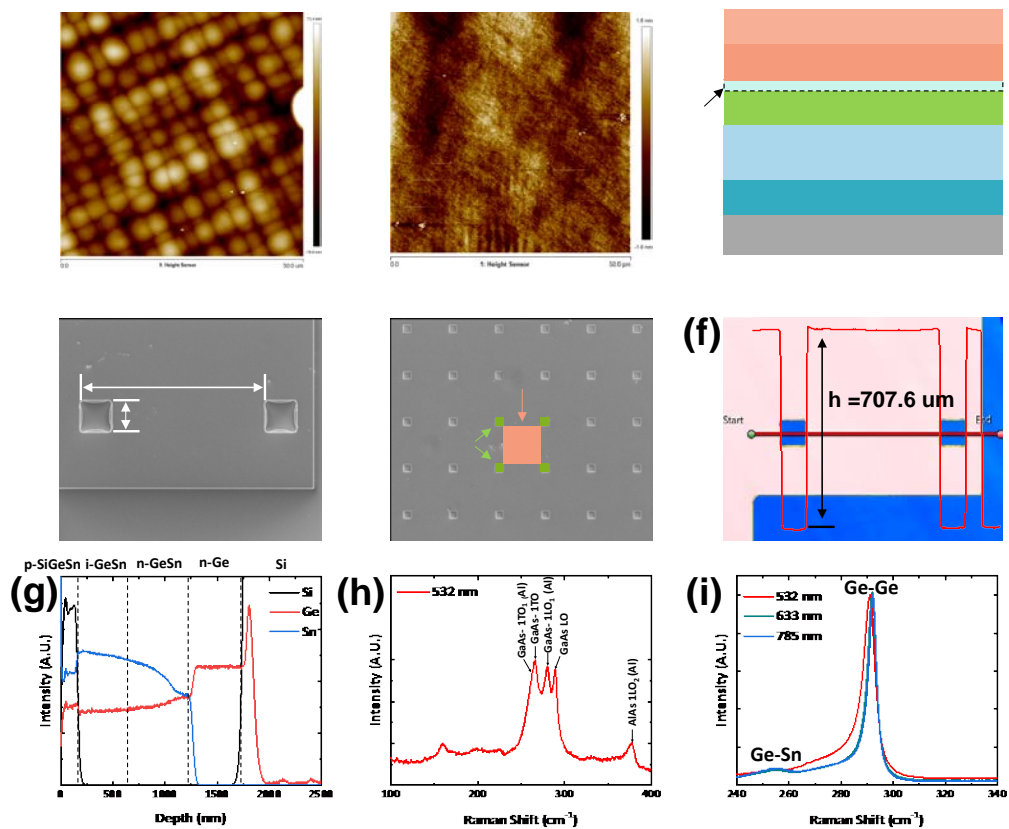

Fig. 2. AlGaAs/GeSn p-i-n heterojunction and materials characterizations (a) AFM roughness measurement of the original GeSn epi substrate; (b) AFM measured roughness of the GeSn epi surface after CMP and HF cleaning; (c) Schematic illustration of the layer structures of the AlGaAs/GeSn heterostructure; (d)(e) Scanning Electron Microscope (SEM) images of the meshed holes on the GaAs/AlGaAs nanomembrane and the seamless integration of the nanomembrane on the GeSn substrate; (f) A plane view of the AlGaAs NM profile with a scanning line detailing step depth across mesh holes; (g) SIMS profile of the original GeSn substrate before CMP; (h) Raman spectra of the grafted AlGaAs NM on the GeSn substrate (the red region in (e)) under the excitation of a 532 nm laser; (i) Raman spectra of the GeSn (The green regions in (e)) under the excitation of 532 nm, 633 nm and 785 nm lasers.

Figures 2(a) and 2(b) illustrate the differences in surface roughness before and after CMP and a 1-minute cleaning with diluted hydrofluoric acid (1:20). The root-mean-square deviation (Rq) across a 50×50 µm² area on the GeSn substrate decreased significantly from 24.6 nm to 0.422 nm, achieving an ultra-smooth surface essential for high-quality grafting. Figures 2(d) and 2(e) display the SEM images of the grafted AlGaAs/GeSn heterostructure. A depth profile across meshed holes, captured using a Filmetrics optical profiler, is shown in Figure 2(f). This analysis revealed an AlGaAs nanomembrane thickness of 707.6 nm atop the GeSn substrate, with a 55 µm separation between the 9×9 µm² meshed holes.

The Raman spectra measured by a Horiba LabRAM HR Evolution Raman spectrometer of the both GaAs/AlGaAs area (the red area in Fig. 2(e)) and exposed GeSn hole area (the green area in Fig. 2(e)) of the AlGaAs/GeSn heterostructure are shown in the Fig. 2(h) and Fig. 2(i), correspondingly. Raman peaks were observed at 263, 268, 280, 289, and 377 cm$^{-1}$ under the excitation of 532 nm laser. The high intensity peak at 268 and 289 cm$^{-1}$ correspond to the first-order transverse optical (1TO) phonon and first-order longitudinal optical (1LO) phonon from the top p$^+$-GaAs layer[48]. Peak at 280 cm$^{-1}$ and 377 cm$^{-1}$ are attributed to the GaAs-like 1LO(1LO$_1$) and AlAs-like 1LO(1LO$_2$) phonons from the n$^-$-AlGaAs layer, respectively[49]. The weak scattering mode observed at 263 cm$^{-1}$ correspond to the GaAs-like 1TO phonon from the AlGaAs layer[50].

To clearly distinguish the Ge-Sn Raman peaks from the GeSn substrate, additional characterizations were conducted using 633 nm and 785 nm laser excitations, which significantly improved the signal-to-noise ratio and allowed for clear identification of Ge-Ge and Ge-Sn vibrational modes[51]. The Ge-Sn peaks around 260 cm$^{-1}$ were clearly observed under 633 nm and 785 nm, which are in relatively good agreement with the reported results[52]. Notably, no Si-Ge peaks were detected under any of the three excitation energies, confirming the effective removal of the SiGeSn cap from the original GeSn substrate through CMP. A prominent Ge-Ge peak at approximately 290 nm was consistent across all examined wavelengths (532 nm, 633 nm, and 785 nm). The Raman spectra from both the AlGaAs and GeSn regions demonstrated monocrystalline characteristics, underscoring the high-quality formation of our AlGaAs/GeSn heterojunction.

To accurately determine the bandgap values of two alloys AlGaAs and GeSn, photoluminescence measurements were performed on both the GaAs/AlGaAs epitaxial layer and the pristine $Ge_{0.853}Sn_{0.147}$ substrate. As depicted in Figures 3(a1) and 3(a2), the photoluminescence spectrum peaks are located at 685.6 nm for the AlGaAs and 2856 nm for the $Ge_{0.853}Sn_{0.147}$ substrate. From these data, the bandgap of AlGaAs is calculated to be approximately 1.81 eV, while that of the pristine $Ge_{0.853}Sn_{0.147}$ substrate is estimated at 0.434 eV.

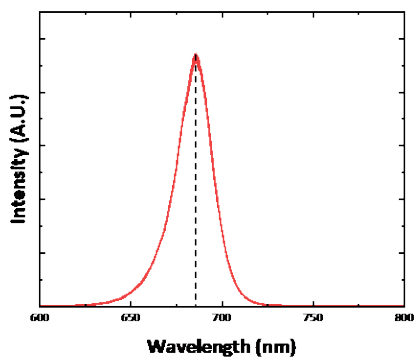
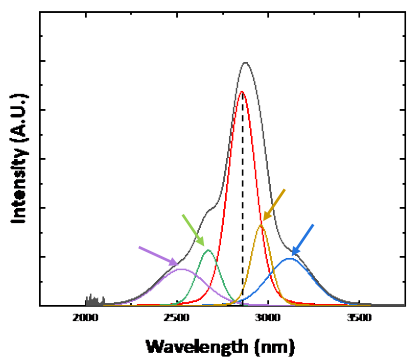

**(b2)**

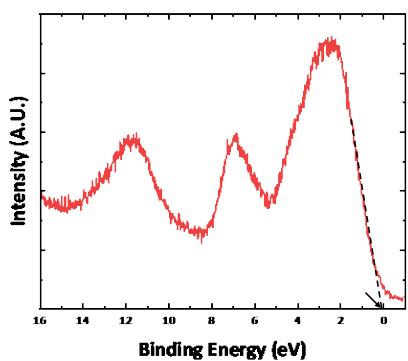
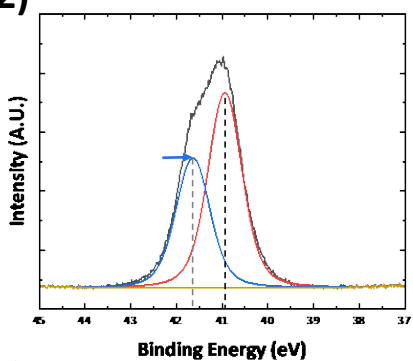

**(c2)**

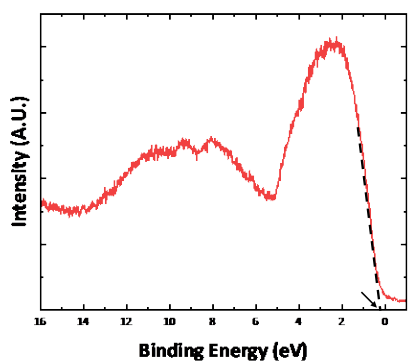
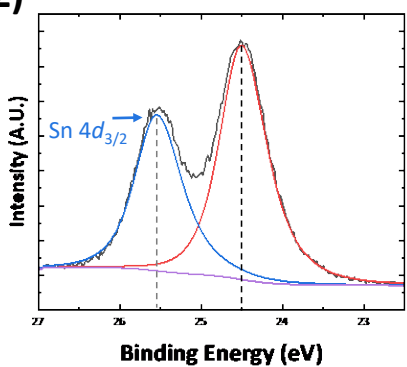

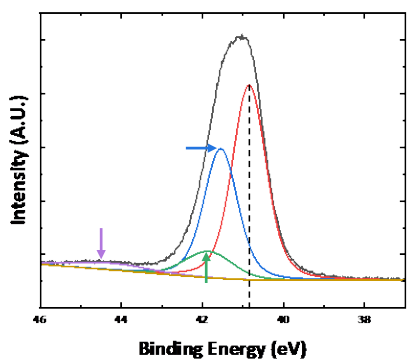
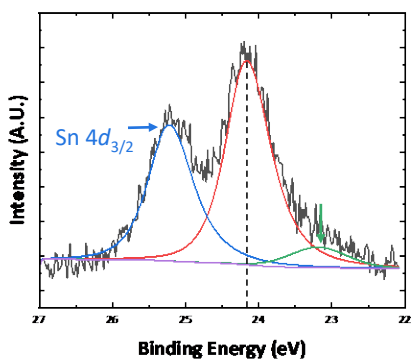

Fig. 3. (a1) The PL spectrum of the AlGaAs NM. (a2) The PL spectrum of the pristine $Ge_{0.853}Sn_{0.147}$ substrate; (b1) The valence band XPS spectrum of the AlGaAs NM with extraction of the valence band maximum (VBM); (b2) The As 3*d* XPS spectrum collected from the AlGaAs NM with extraction of the As 3*d* peak positions; (c1) The valence band XPS spectrum of the pristine GeSn substrate with extraction of the VBM; (c2) The Sn 4*d* XPS spectrum obtained from the pristine GeSn substrate with extraction of the Sn 4*d* peak positions; (d1) The As 3*d* XPS spectrum of the grafted AlGaAs/ $Ge_{0.853}Sn_{0.147}$ sample with extraction of the As 3*d* peak positions; (d2) The Sn 4*d* XPS spectrum of the grafted AlGaAs/ $Ge_{0.853}Sn_{0.147}$ heterostructure sample with extraction of the Sn 4*d* peak positions.

The Valence Band Offset (VBO) between AlGaAs and $Ge_{0.853}Sn_{0.147}$ was determined using the method developed by Kraut et al. in 1980[46]. Measurements were performed on both the pristine GeSn substrate and the AlGaAs nanomembrane (NM). For the GeSn substrate, the Sn $4d_{5/2}$ core level peak were recorded at 24.51 eV (Fig. 3(c2))[53]. The valence band maximum (VBM) was determined at 0.22 eV (Fig. 3(c1)). For the AlGaAs NM, the As $3d_{5/2}$ core level peak were measured at 40.94 eV (Fig. 3(b2))[54]. The VBM of AlGaAs was at 0.16 eV (Fig. 3(b1)). In the grafted AlGaAs/GeSn structure, the Sn $4d_{5/2}$ core level and As $3d_{5/2}$ core level peaks were observed at 24.17 eV (Fig. 3(d2)) and 40.85 eV (Fig. 3(d1)), respectively. Weak signals of $As_2O_3$ and $AsO_x$ were observed as shown in the Fig. 3(d1)[55]. Following Kraut et al.'s method, using As $3d_{5/2}$ and Sn $4d_{5/2}$, the VBO is calculated using the equation[46]:

$$\Delta E_V = (E_{CL}^Y - E_V^Y) - (E_{CL}^X - E_V^X) - \Delta E_{CL}, \tag{1}$$

where $\Delta E_V = E_V^X - E_V^Y$ represents the VBO, $\Delta E_{CL} = E_{CL}^Y(i) - E_{CL}^X(i)$ is the difference in binding energy between core level peaks of semiconductors X and Y at the interface, $E_{CL}^Y - E_V^Y$ is the binding energy difference between the core level peak and VBM in semiconductor Y, and $E_{CL}^X - E_V^X$ is the binding energy difference between the core level peak and VBM in semiconductor X. In this study, GeSn represents semiconductor X and AlGaAs the semiconductor Y. The corresponding VBO value was calculated to be -0.19 eV. Combining this result with the measured $Ge_{0.853}Sn_{0.147}$ bandgap value (0.434 eV) and AlGaAs (1.81 eV), the Conduction Band Offset (CBO) value between AlGaAs and $Ge_{0.853}Sn_{0.147}$ was calculated to be +1.186 eV.

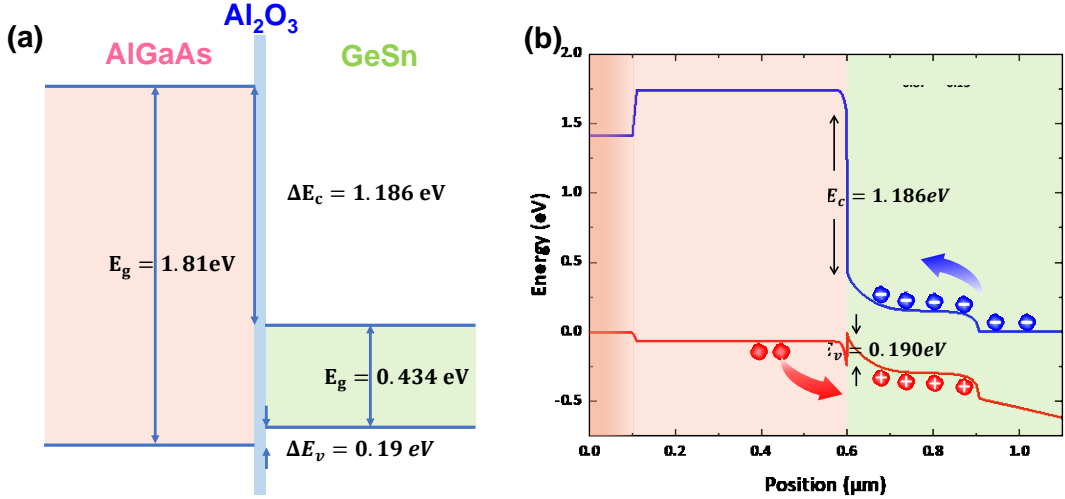

Fig. 4. (a) Schematic band alignment of the grafted AlGaAs/GeSn heterostructure, constructed based on core level peak results from XPS measurements; (b) Equilibrium band diagram of the heterostructure, illustrating the confinement properties of electrons and holes.

The band diagram of the grafted monocrystalline $Al_{0.3}Ga_{0.7}As$/ $Ge_{0.853}Sn_{0.147}$ heterostructure was reconstructed based on the results obtained from the above method, as shown in the Fig. 4(a). The AlGaAs and GeSn layers form a type-I band alignment, creating a conduction band offset of 1.186 eV. This offset, which functions as an electron barrier, is notably higher than those reported for $Si_{0.03}Ge_{0.89}Sn_{0.08}/Ge_{0.89}Sn_{0.11}$ DHS (114 meV)[28], $Ge_{0.925}Sn_{0.075}/Ge_{0.906}Sn_{0.094}/Ge_{0.891}Sn_{0.109}$ DHS (10 meV)[56], and $Ge/Ge_{0.92}Sn_{0.08}/Ge$ multiple quantum well structures (18 meV)[57]. The Silvaco simulated band structure of the grafted AlGaAs/GeSn/GeSn p-i-n double heterostructure, using band parameters derived from the XPS results, is depicted in Figure 4(b). In this structure, holes are injected from the p$^+$-GaAs/ p$^-$-AlGaAs cap layer and are confined by the grading n$^+$-GeSn layer, while electrons are injected from the n-GeSn side and confined by the robust 1.186 eV CBO. It is expected that the improved charge confinement can improve the operation temperature of GeSn based lasers.

## 4. Conclusion

In this study, we conducted a comprehensive analysis of the band offset in the monocrystalline grafted Al$_{0.3}$Ga$_{0.7}$As / Ge$_{0.853}$Sn$_{0.147}$ heterostructure using XPS. The bandgap for Ge$_{0.853}$Sn$_{0.147}$ was determined to be 0.434 eV, while that for Al$_{0.3}$Ga$_{0.7}$As was found to be 1.81 eV attracted from photoluminescence. The valence band offset and conduction band offset values, calculated using core level peaks and valence band maximum, revealed a Type-I band alignment. This insight underscores the potential of the grafting technique in advancing the development of GeSn heterostructure-based devices and could significantly impact the future applications of such technologies in development of GeSn laser devices with high operation temperature.

**Acknowledgements**

This work was supported by Air Force Office of Scientific Research (AFOSR) under Grant No. FA9550-19-1-0102 and National Science Foundation (NSF) under Award No. 2235443.

**CRediT authorship contribution statement**

**Yang Liu:** Writing - review & editing, Writing - original draft, Visualization, Validation, Methodology, Investigation, Formal analysis, Data curation, Conceptualization. **Jiarui Gong:** Writing - review & editing, Writing - original draft, Visualization, Validation, Methodology, Investigation, Formal analysis, Data curation, Conceptualization. **Sudip Acharya:** Visualization, Methodology, Investigation, Data curation. **Yiran Li:** Methodology. **Alireza Abrand:** Methodology. **Justin M. Rudie:** Methodology, Writing - review & editing. **Jie Zhou:** Methodology, Writing - review & editing. **Yi Lu:** Methodology, Writing - review & editing. **Daniel Vincent:** Methodology. **Samuel Haessly:** Methodology. **Tsung-Han Tsai:** Methodology. **Parsian K. Mohseni:** Resources, Funding acquisition. **Shui-Qing Yu:** Writing - review & editing, Writing - original draft, Validation, Supervision, Resources, Methodology, Project administration, Funding acquisition. **Zhenqiang Ma:** Concept design, Writing - review & editing, Writing - original draft, Validation, Supervision, Resources, Project administration, Methodology, Funding acquisition, Formal analysis, Conceptualization.